# Solutions to the Cosmic Initial Entropy Problem without Equilibrium Initial Conditions

Vihan M. Patel [1] and Charles H. Lineweaver [1,2,3,*]

[1] Research School of Astronomy and Astrophysics, Australian National University, Canberra 2600, Australia; vihan.i.patel@gmail.com
[2] Planetary Science Institute, Australian National University, Canberra 2600, Australia
[3] Research School of Earth Sciences, Australian National University, Canberra 2600, Australia
[*] Correspondence: charley.lineweaver@anu.edu.au

**Abstract:** The entropy of the observable universe is increasing. Thus, at earlier times the entropy was lower. However, the cosmic microwave background radiation reveals an apparently high entropy universe close to thermal and chemical equilibrium. A two-part solution to this cosmic initial entropy problem is proposed. Following Penrose, we argue that the evenly distributed matter of the early universe is equivalent to low gravitational entropy. There are two competing explanations for how this initial low gravitational entropy comes about. (1) Inflation and baryogenesis produce a virtually homogeneous distribution of matter with a low gravitational entropy. (2) Dissatisfied with explaining a low gravitational entropy as the product of a 'special' scalar field, some theorists argue (following Boltzmann) for a "more natural" initial condition in which the entire universe is in an initial equilibrium state of maximum entropy. In this equilibrium model, our observable universe is an unusual low entropy fluctuation embedded in a high entropy universe. The anthropic principle and the fluctuation theorem suggest that this low entropy region should be as small as possible and have as large an entropy as possible, consistent with our existence. However, our low entropy universe is much larger than needed to produce observers, and we see no evidence for an embedding in a higher entropy background. The initial conditions of inflationary models are as natural as the equilibrium background favored by many theorists.

## 1. The Entropy of the Universe, the Second Law, the Past Hypothesis, and the Cosmic Initial Entropy Problem

As stars shine and black holes accrete, the entropy of the universe goes up [1]. Galaxies with shining stars and accreting black holes are distributed relatively homogeneously across the universe. This observed homogeneity of the universe on large scales implies an approximately zero net flow of entropy between volumes larger than a scale of a few hundred million light years. These large representative volumes of the universe are effectively closed. Thus, the second law of thermodynamics applies to the entire universe, and its entropy does not decrease: $dS \geq 0$ [2–4]. In particular, the entropy of the observable universe $S_{uni}$ (defined as the entropy of the comoving volume of our current particle horizon) does not decrease [5–7]. Thus, the entropy of the universe was smaller in the past [8,9] and will be larger in the future. This requirement of low entropy conditions in the early universe is often called the 'past hypothesis' [10–13]. The increasing entropy of the universe will eventually approach a maximum entropy state: $S_{max}$ ([1,14,15], but see 16 for an opposing view).

The second law and the past hypothesis predict that the early universe was at low entropy. However, the photons in the cosmic microwave background (CMB) have temperature deviations of only $\frac{\Delta T}{T} \sim 10^{-5}$ [17] around their average blackbody temperature of 2.7 K [18]. The entropy of a given comoving volume of blackbody photons in an expanding universe (including deSitter universes) remains constant [1,19] and corresponds to maximal entropy for the photons. Thus, the cosmic microwave background radiation is remarkably close to an equilibrium blackbody spectrum and reveals an apparently high entropy universe close to thermal and chemical equilibrium. The cosmic initial entropy problem is illustrated in Figure 1a. Solutions to the cosmic initial entropy problem (IEP) need to explain both the initial low entropy of the universe (required by the second law), and the apparent high entropy of the observed CMB.

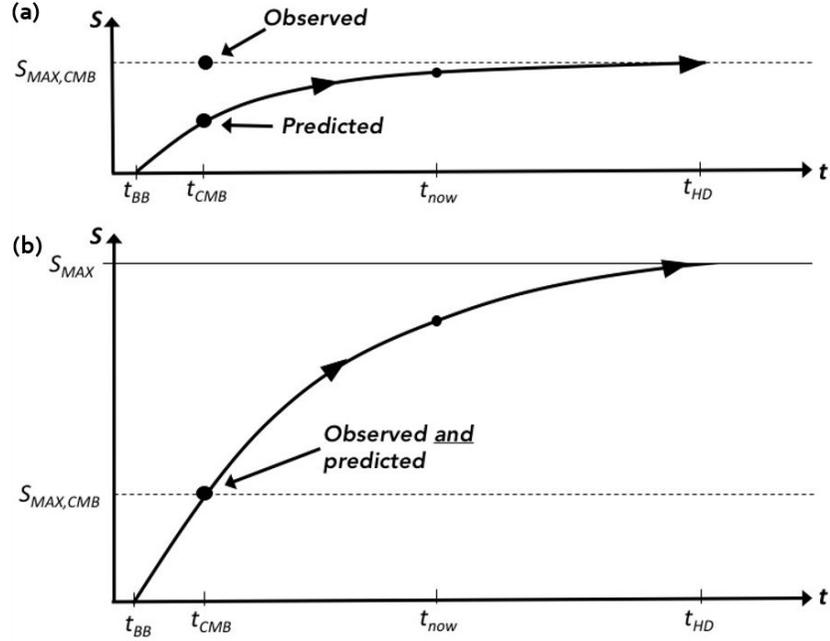

**Figure 1.** The Initial Entropy Problem. (**a**) The second law and the past hypothesis make a low entropy prediction for the early universe. However, observations of the cosmic microwave background (CMB) show a universe at thermal and chemical equilibrium, i.e., maximum entropy. The problem is resolved in (**b**) when we include the low gravitational entropy of the homogeneous distribution of matter in the early universe and define a new maximum entropy that includes gravitational entropy: $S_{MAX} = S_{MAX,grav} + S_{MAX,CMB}$. Also, we require $S_{MAX,grav} \gg S_{MAX,CMB}$. Thus, the inclusion of gravitational entropy resolves the discrepancy between our expectations of a low entropy beginning and the observed high entropy of the CMB.

## 2. Gravitational entropy and Penrose's Weyl Curvature Hypothesis

### 2.1. Kinetically vs. Gravitationally-Dominated Systems

In kinetically-dominated systems, any concentrated particles (e.g., perfume in a bottle) diffuse until the particles are distributed relatively homogeneously. A homogeneous distribution corresponds to a state of thermal equilibrium and maximum entropy. Penrose [6,19] has suggested that gravitationally-dominated systems behave in the opposite way: a smooth distribution of matter corresponds to minimal gravitational entropy. In the panels of Figure 2b, starting from a close-to-homogeneous distribution, the entropy increases with gravitational collapse to black holes [14]. The highest gravitational entropies—as well as the largest contributions to the entropy of the universe— are in the supermassive black holes at the centre of many galaxies [4,7]. The eventual Hawking evaporation of the black holes [21] increases the entropy even further [22–24]. The gravitational entropy ($S_{grav}$) of this progression from a smooth distribution of matter, to black holes and then to the photons from black hole evaporation, cannot yet be expressed and quantified in an equation of the form $S_{grav} = f(P(k,t))$, where $P(k,t)$ is the time-dependent power spectrum of large-scale structure. In order for $S_{grav}$ to solve the initial entropy problem, gravitational collapse and the resulting change in $S_{grav}$ has to be the source of all entropy increase since the end of inflation.

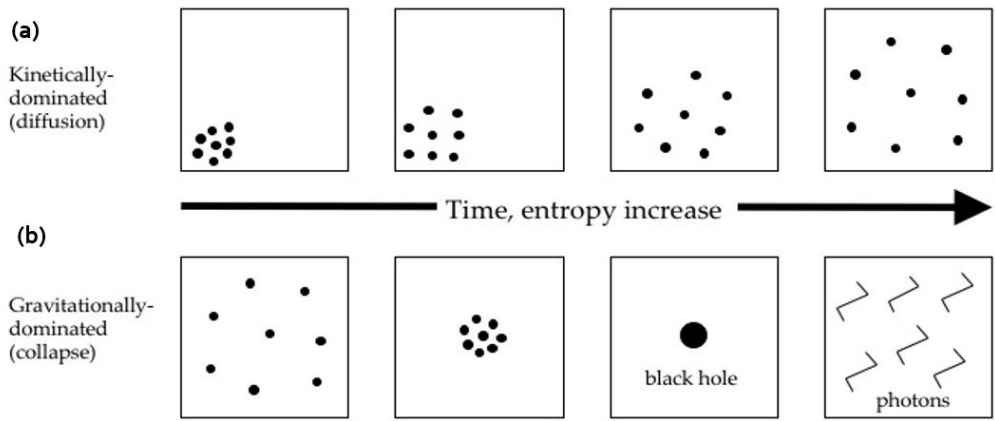

**Figure 2.** Comparison of the entropic evolution of (**a**) a kinetically-dominated system undergoing diffusion and (**b**) a gravitationally-dominated system undergoing collapse and evaporation via Hawking radiation. The relationship between gravitational collapse and the increase in entropy is not well established, but to solve the initial entropy problem and to be consistent with the high entropy of black holes, it must be as sketched here.

*2.2. Penrose's Weyl Curvature Hypothesis*

Penrose's Weyl Curvature Hypothesis identifies the amount of gravitational curvature with the value of the Weyl conformal tensor. The off-diagonal elements of the Weyl conformal tensor are constrained to zero or near zero at initial singularities where $t \rightarrow 0$ [20]. This would constrain the initial gravitational entropy $S$ of the universe to be low:

> The Weyl curvature vanishes…at the initial singularity and is unconstrained, no doubt diverging wildly to infinity, at final singularities.
>
> —Penrose [6] (p. 767)

The thermodynamics of gravitational fields remains contentious [25,26]. However, a fundamental relationship between entropy, gravity, and quantum mechanics may exist ([6], p. 692). Gravitational entropy offers a solution to the initial entropy problem, with low entropy states being defined as close-to-homogeneous distributions of matter, able to gravitationally collapse and increase their entropy (Figure 2b).

**3. Inflation Produces Low Entropy Initial Conditions**

The inflationary paradigm [27–29] solves the horizon, flatness, and monopole problems of cosmology, and is often included in solutions to the cosmic initial entropy problem [16,30,31]. Cosmological observations are consistent with inflation and therefore support it as the mechanism responsible for the initially low entropy of the observable universe and presumably of the entire inflationary bubble.

In inflationary models, during reheating, unclumpable false vacuum energy is dumped into the universe as a virtually homogeneous distribution of relativistic particles. Baryogenesis and the expansion and cooling of the universe then produce an homogeneous distribution of clumpable matter, which provides the low gravitational entropy state. In this model, the low initial entropy of the universe comes from the even lower initial entropy of a scalar field on the flat part of the inflaton potential (Figure 3a).

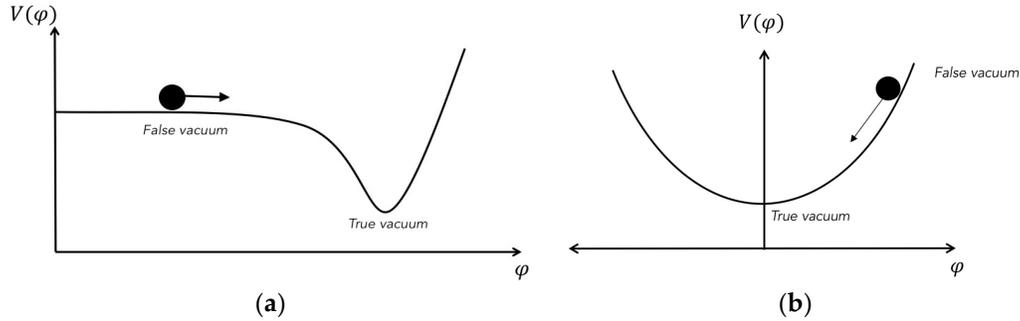

**Figure 3.** The two panels represent competing models for the shape of the inflaton potential: (**a**) 'slow roll' inflation; (**b**) chaotic inflation. In both, cosmic inflation begins when the scalar field $\varphi$ rolls down its potential from a false vacuum to a true vacuum, creating an expansion of space of at least 60 e-foldings either at the GUT scale (~$10^{-35}$ seconds after the big bang) or at the Planck scale (~$10^{-43}$ seconds after the big bang). The field oscillates around its minimum true vacuum state and interacts with other fields during a reheating phase in which particles are produced [32,33].

The proto-inflationary state has its energy density stored in a single degree of freedom—the unstable potential of the scalar field: $V(\varphi)$. Carroll and Chen ([16], p. 12) estimate the entropy of $\varphi$ at $S \sim 10^{12}$. The difficulty of explaining the 'unnatural', special state of $\varphi$ motivates the universe-from-equilibrium models.

## 4. Boltzmann's Anthropic Hypothesis: Low Entropy Fluctuation in a Maximum Entropy Background

Boltzmann first proposed that the low entropy of the universe was a random fluctuation from a maximal entropy state [34,35]. This is an example of Poincare recurrence [36] and can be thought of as the time reversal of Figure 2a: the random velocities of a diffuse gas in a room conspire to move all the particles into a single corner. This is highly unlikely, but given an infinite amount of time, it will occur again and again. The Boltzmann fluctuation model is the most general description of what recent models propose—an equilibrium state that moves to low entropy via a stochastic fluctuation.

The fluctuation theorem [37] quantifies such fluctuations away from maximal entropy and yields the result that small fluctuations are exponentially more likely than large fluctuations (Figure 4b). A potentially spatially infinite universe and anthropic considerations suggest that a low entropy fluctuation sufficient for our existence is a possibility, but importantly, that such a fluctuation would be minimally sufficient for our existence.

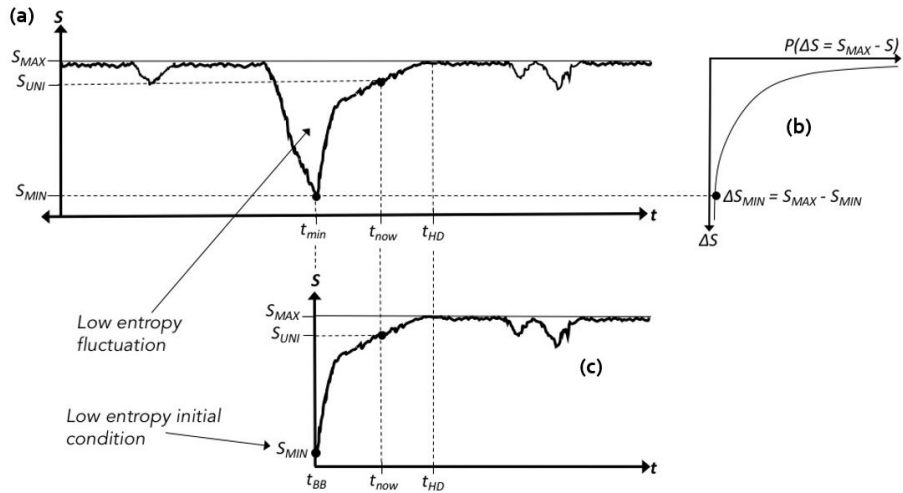

**Figure 4.** Two different solutions to the initial entropy problem. (**a**) Boltzmann's idea that our universe is a low entropy fluctuation surrounded by a universe in equilibrium at maximum entropy. The spectrum of the fluctuations away from $S_{MAX}$ should agree with the fluctuation theorem [37]; (**b**) The probability P of fluctuations of amplitude $\Delta S = S_{MAX} - S$. Small fluctuations away from equilibrium are exponentially more likely than large fluctuations; (**c**) The low entropy initial condition is the result of inflation and the homogeneous distribution of matter (i.e., a state of low gravitational entropy) that it produces. Time before the big bang is not part of this non-equilibrium, inflation-only model.

*4.1. Problems with Boltzmann's Hypothesis*

One galaxy or even only one star with planets embedded in a universe of cold blackbody photons would presumably be enough to produce observers. However, in our observable universe, we see hundreds of billions of galaxies each with hundreds of billions of stars. We observe a far greater number of independent low entropy structures than necessitated by the anthropic principle [38,39].

In 1963, before the discovery of the CMB and before Penrose's postulate that an even distribution of matter corresponds to low gravitational entropy, Feynman made this same point:

> …from the hypothesis that the world is a fluctuation, all of the predictions are that if we look at a part of the world we have never seen before, we will find it mixed up, and not like the piece we looked at. If our order was due to a fluctuation, we would not expect order anywhere but where we have just noticed it.
>
> —Feynman [40] (lecture 46)

Davies also makes this point:

> The fact that we inhabit at least a Hubble volume of low entropy must be counted as strong evidence against Boltzmann's hypothesis.
>
> —Davies [41] (p. 9)

Figure 5 is a conformal diagram of Boltzmann's Hypothesis. Our increasing particle horizon should reveal a high entropy background different from the CMB fluctuations consistent (through the Sachs–Wolfe effect [42] and baryonic acoustic oscillations [43]) with the relatively smooth distribution of matter that evolves into the cosmological large scale structure we see around us today. If Boltzmann were correct, as our particle horizon has increased, we should have been able to see the maximum entropy universe that did not contribute to our existence. It would be different from the low entropy universe that did contribute to our existence.

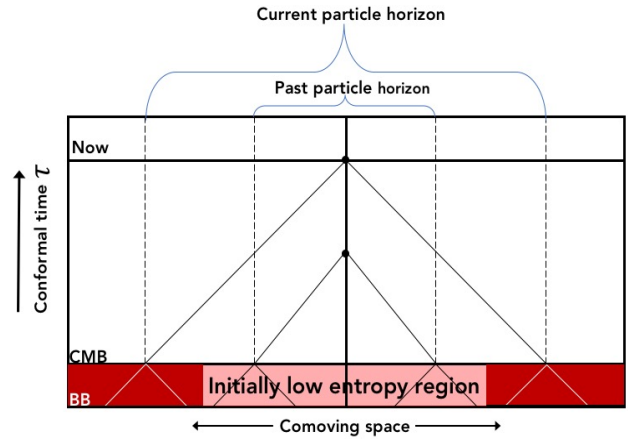

**Figure 5.** Observational evidence against Boltzmann's Hypothesis. As time goes by, our particle horizon increases in size, i.e., the current particle horizon is larger than the past particle horizon. Therefore, we are able to see parts of the universe that we have not been able to see before. We are able to see parts of the universe that did not have to be at low entropy for us to be here. If our universe began at low entropy (pink) embedded in a maximal entropy universe in equilibrium (red), then anisotropies in the CMB would not be at the right level to represent density fluctuations that grow into large scale structure, but would represent the evaporated photons from black holes (right of Figure 2b).

If we take the low entropy bubble necessitated by anthropic constraints to be the size of the solar system, or (more conservatively) the galaxy, then we should not observe low-entropy structures elsewhere in the universe. Yet, the increasing particle horizon volume reveals new, anthropically irrelevant galaxies and the low gravitational entropy of the small-density-contrast seeds of galaxies in the CMB. These newly visible low entropy regions played no role in our being here.

*4.2. Boltzmann Brains: How Small Can the Low Entropy Region Be and Still Produce Observers?*

Anthropic constraints placed on the Boltzmann fluctuation model suggest that rather than a low entropy fluctuation that allows for the evolution of a conscious observer 'from scratch' (e.g., the development of a star, around which a planet develops the requisite conditions for life, which then evolves into conscious entities), one should expect a fluctuation into conscious entities directly.

Since low entropy (as a result of random fluctuations) is at a premium, Albrecht & Sorbo think we should expect the size of the low entropy region to be as small as possible [44]. If biological evolution is not required to produce an observer, then we should expect minimalistic observers in the form of statistical fluctuations. Albrecht & Sorbo call this the 'Boltzmann brain paradox' [44]. Minimal fluctuations that directly produce self-aware observers (floating 'brains' in a maximal entropy background) should be much more likely than the evolution of observers within a local low entropy fluctuation the size of the observable universe. There should be far more Boltzmann brains floating in an equilibrium than observed. The fact that we observe no such Boltzmann brains suggests that either natural selection and the ratcheting of biological evolution is much more efficient than random fluctuations at producing observers, or that the whole idea of our universe being an anthropic low entropy island is wrong.

## 5. Which Initial Condition is More 'Natural', Inflation or Equilibrium?

A low entropy state corresponds to a system in a macrostate with a relatively small number of microstates; the system is in an improbably small region of phase space. Discussions of the initial entropy problem often attempt to provide 'natural' solutions to our supposed thermodynamically 'unnatural' low entropy origins. Equilibrium states are assumed to be the most probable states in phase space and are associated with naturalness under the standard Boltzmann statistics. Unhappy with the unnaturalness of inflationary conditions, equilibrium models or 'Universe from chaos' models similar to Boltzmann's original hypothesis have been proposed [16,30,31].

> The goal I am pursuing is to find cosmological scenarios in which the Past Hypothesis is predicted by the dynamics, not merely assumed.
>
> —Carroll [13]

In equilibrium models, the universe should initially exist in an equilibrium state. Equilibrium is assumed as an initial boundary condition. The universe should also be able to produce universes resembling our own via some dynamical mechanism—usually in the form of quantum fluctuations. These fluctuations can be arbitrarily rare, but anthropic reasoning allows us to consider any fluctuation with a non-zero probability of occurrence, sufficient for solving the initial entropy problem.

*What is Wrong with Equilibrium as an Initial Condition?*

The application of statistical mechanics to the notion of a cosmological initial condition is problematic. Systems as a whole are expected and observed to evolve into states of maximum entropy over time. Dynamical processes increase the entropy until equilibrium is reached. In phase space, appropriately distinguishable macrostates are 'coarse grained' into distinct volumes with different sizes depending on the number of corresponding microstates (see [6], p. 691, Figure 27.2 demonstrating this evolution visually). A system that begins in a macrostate with a small coarse grained volume is expected to move into larger volumes. This is by virtue of the ergodic assumption that every microstate is accessible and that the macrostates with the largest volumes are the ones systems most probably end up in. However, 'end up in' and 'begin in' are different. Natural and stable end states are not necessarily initial states.

A true 'initial condition' is not expected to have evolved towards larger volumes since there are not prior states from which it began. If we are considering 'initial condition space' as the totality of different ways a universe can begin, then there is no reason to believe that the largest volume is most probably the initial condition because there is no evolving system sampling ever-increasing volumes, to ensure a uniform distribution of initial states over the "allowed" space. The force of the second law and its drive towards equilibrium cannot exist without evolution from one state to another, and the question of the most probable initial condition cannot be determined by the statistical mechanics of the second law because these statistics determine how a system will end up from any point, not how the system should be at an initial point.

Proponents of equilibrium cosmology argue that the initial condition should not be assumed a priori as 'special' but rather should be 'generic'. Existing physics should be sufficient in setting up the initial conditions for our universe from a generic state ([6], p. 756). However, a generic state is not as well-defined a concept in the context of initial conditions, as it is in the context of thermodynamics. There is no known probability distribution for initial conditions, and no pre-initial-condition mechanism for a system to diffuse and attain a priori the most probable a posteriori macrostate.

Carroll and Chen ([16], p. 6) justify the requirement for high entropy initial conditions by invoking an argument for time-symmetry. They argue that we should expect 'natural' initial conditions to resemble 'natural' final conditions if we accept time-symmetric physics. In this context a low entropy initial condition is not 'natural' because the final state of a recollapsing universe would not be expected to 'deflate' (at least in the inflationary scenario) [30,45]. Carroll and Chen are making a positive claim as to the origin of the thermodynamic arrow of time and its relation to other observed arrows (e.g., cosmological, psychological, etc.), namely that time invariant microscopic physics supersedes the second law. We opt to accept the second law at face value and assume that it describes a fundamental aspect of reality: time asymmetric boundary conditions (i.e., low initial entropy, high final entropy). Time-asymmetric boundary conditions are acceptable under the current understanding (or lack thereof) of initial condition space and the limitations of Boltzmann statistics. It is not necessarily more natural to impose time-symmetric boundary conditions.

We suggest a low entropy initial condition is favored over a high entropy one by virtue of the second law and the past hypothesis. Additional assumptions must be made in order to generate a low entropy state from an equilibrium initial condition (i.e., a dynamical mechanism from which we generate a low entropy state from a high entropy one).

Testing inflationary cosmology is a work in progress [46–49]. Inflation remains the consensus early-universe-add-on to general relativity and as such is attractive to resolve the initial entropy problem. We do not attempt to provide a dynamical solution to the 'special' initial state of the scalar field since dynamical solutions are generally founded on the basis of equilibrium cosmology, and because the appropriate physics describing the early universe close to the Planck time does not exist.

## 6. Summary and Conclusions

Solutions to the initial entropy problem rely on Penrose's Weyl Curvature Hypothesis [20] to explain the low gravitational entropy of the early universe. Most authors (with the exception of Penrose) accept inflationary models as the mechanism for producing this early low gravitational entropy. We argue that inflationary models can be accepted as initial conditions until we know more about Planck-scale physics. Unhappy with the unnaturalness of inflationary conditions, equilibrium models or 'Universe from chaos' models similar to Boltzmann's original hypothesis have been proposed (e.g., [16,30,31]). These models are based on the idea that a low entropy fluctuation from an equilibrium background is a natural solution to explaining low initial entropy. These models often assume time-symmetric boundary conditions which seem to us no more compelling than time-asymmetric boundary conditions.

Penrose's rejection of inflation prevents him from accepting the most strongly supported model of the early universe—one that produces the initial low gravitational entropy that he has championed. We argue that inflation, combined with expansion, cooling, and baryogenesis, produces a relatively smooth distribution of matter that is equivalent to the low gravitational entropy needed to explain the initial entropy problem. Thus, in addition to the flatness, horizon, and monopole problems, inflation solves the cosmic initial entropy problem.

**Acknowledgments:** C.H.L. thanks Andy Albrecht and Paul Davies for insightful discussions.

**Author Contributions:** V.M.P. and C.H.L. jointly conceived the ideas, made the figures and wrote the paper.

**Conflicts of Interest:** The authors declare no conflict of interest.